\documentclass[preprint,proceedings]{rmaa}

\suppressfulladdresses 

\usepackage{paralist}

\SetYear{2005}
\SetConfTitle{Latin American Regional IAU Meeting, Pucon, Dec 2005}

\title{High Resolution Spectra of Novae and the Quadratic Zeeman Effect} 

\author{
  Robert Williams\altaffilmark{1} 
  and Elena Mason\altaffilmark{2}}

\altaffiltext{1}{Space Telescope Science Institute, Baltimore, MD, USA}

\altaffiltext{2}{European Southern Observatory, Santiago, CL}

\shortauthor{Williams \& Mason}
\shorttitle{RevMexAA(SC) Classical Novae \& Quadratic Zeeman Effect}

\listofauthors{Robert Williams \& Elena Mason}
\indexauthor{Williams, Robert}
\indexauthor{Mason, Elena}

\abstract{High  resolution  spectra  of  novae after  outburst  reveal
distinctive characteristics in the line profiles and intensities.  The
higher Balmer  lines are often broader  than the lower  members of the
series,  and  the relative  profiles  and  intensities  of the  [O~I]
$\lambda\lambda$6300,  6364  doublet differ  from  normal values.   We
suggest these  features may be  caused by the Quadratic  Zeeman Effect
from  magnetic fields  exceeding  B=10$^6$ gauss.  Taken together  the
emission and absorption lines point to multiple origins for the ejecta
on both the erupting white dwarf and the cool secondary star. }

\addkeyword{Stars: Classical novae}
\addkeyword{Stars: Polars}

\begin{document}
\maketitle

\section{Introduction}
\label{sec:intro}

Emission  line spectroscopy  has  been done  predominantly  at low  to
moderate  resolutions in recent  years because  many of  the important
diagnostics require only relative line intensities.  Larger telescopes
with  more efficient  instruments now  allow high  spectral resolution
data to  be obtained  for many objects,  and these  provide additional
information  that   is  useful  for  spectral   analysis.   Among  the
advantages of high spectral resolution  are the (1)~detection of faint
features and  narrow lines that escape detection  at lower resolution,
(2)~resolution of blends into individual lines, and (3)~measurement of
line  profiles, which  provide kinematic  information on  the emitting
gas.

 The larger number of lines that are detected in deep, high resolution
 spectra,  many of which  are very  faint and  have not  been observed
 previously,  make correct  line identification  a  major undertaking.
 Electronic databases of atomic  transitions now exist which make this
 task perfectly suited to  computer algorithms.  A line identification
 algorithm named  EMILI has  been developed by  Sharpee et~al.\ (2003)
 that  is largely  automatic and  which  is based  upon the  extensive
 compilation of transitions by van  Hoof in the v2.05 Atomic Line List
 (http://www.pa.uky.edu/$\sim$peter/newpage/).    Continued   work  is
 underway to improve EMILI by  refining the criteria by which the line
 identifications  are  made, with  the  goal  of eventually  including
 molecular  species  in  the   lookup  database.   Since  proper  line
 identification is the essential  first phase of all spectral analysis
 it   is   important  for   high   signal-to-noise,  high   resolution
 spectroscopy that these efforts be carried forward.

\section{Novae spectra at high resolution}
\label{sec:hr}

\begin{figure*}[!p]
\centering
  \includegraphics[angle=-90,width=12cm]{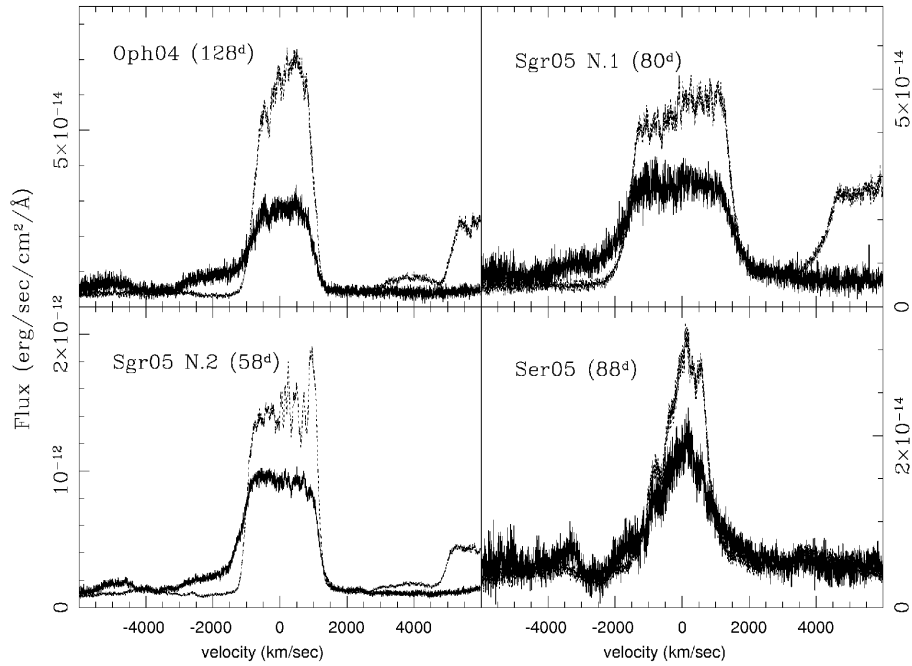}
  \caption{A comparison of H$\beta$ (dotted line) and H$\delta$ (solid
  line)  emission  line  profiles   for  four  novae  in  our  survey.
  H$\delta$,  the  less intense  of  the  two  lines, is  broader  and
  slightly  more  blue-shifted than  H$\beta$,  which  is  one of  the
  important characteristics  of the  Quadratic  Zeeman effect for H (Preston
  1970).  The curves are labeled with the epoch of observation in days
  after outburst.}
  \label{fig:bl}
\end{figure*}

Two important goals of  high resolution spectroscopy are the detection
of  features  that  are  not  observed at  lower  resolution  and  the
measurement of line profiles.  We have collaborated on a spectroscopic
survey of southern postoutburst novae for the past two years using the
ESO  fiber optics echelle  spectrograph FEROS  (R=48,000) on  the 2.2m
telescope, and certain key features of novae spectra have emerged that
were  not  apparent from  the  spectra  of  previous lower  resolution
studies such as the CTIO survey (Williams et~al.\ 1991). These include
(1)~the presence of Na~I~D absorption shortly after maximum in almost
every nova, (2)~the systematic shift  of emission lines of H and He to
more positive radial velocities than those of the forbidden lines, 
(3)~a large fraction  of novae that have progressively  larger line widths
for the higher lines in the Balmer series, and (4)~clearer evidence of
the differences in the line profiles and relative intensities of [O~I]
$\lambda$6300 and  $\lambda$6364, with $\lambda$6364  always being the
broader line.  The first two of these characteristics can be accounted
for straightforwardly by the  ionization and kinematics of the ejected
gas, however the cause of the  last two are less clear.  Optical depth
effects can  explain both the relative differences  between the Balmer
lines and  the [O~I] doublet  line widths, however  in this  case the
lower Balmer  lines and [O~I]  $\lambda$6300 should be  the broader of
the  lines.  But,  the  opposite is  observed,  as is  shown in  Figs.~\ref{fig:bl} 
and~\ref{fig:oi}.   These unusual characteristics are not
observed in  other emission-line objects  and are probably due  to the
unusual and  extreme conditions associated with the  outburst. Table~1
lists some relevant line widths  measured from selected spectra in our
database.  We suggest  here that the differing line  widths may be the
result  of a  breakdown  in  normal LS-coupling  caused  by very  high
electromagnetic field strengths.

\section{The Zeeman effect}
\label{sec:zeeman}
Strong  magnetic or electric  fields change  bound energy  levels such
that  normal relationships governing  transitions are  highly modified
(Herzberg  1945).  The  presence of  forbidden lines  in  the spectra,
which are collisionally suppressed  at high densities, makes the Stark
effect  a less  likely influence  on the  spectral features  than high
magnetic  fields because  the  Stark effect  requires high  densities.
Furthermore, it is already demonstrated that the class of CVs known as
polars,  of  which  the   classical  novae  V1500  Cygni/1975  and  DQ
Herculis/1934 are  examples, have  strong magnetic fields  that exceed
B=10$^6$ gauss (Kaluzny \&  Chlebowski 1988; Warner 1995).  The Zeeman
effect is therefore a plausible  explanation for some of the anomalous
line behavior observed in postoutburst novae.

The  normal   Zeeman  effect  produces  line   broadening  from  level
splitting, by  removing the degeneracy  of magnetic substates,  but it
does not greatly modify  the intensities of entire multiplets relative
to each other.  On the  other hand, the Quadratic Zeeman effect (QZE),
which dominates the normal Zeeman effect for field strengths exceeding
B=10$^6$ gauss (Jenkins \& Segre 1939; Preston 1970; Garstang \& Kemic
1974) results  in much  broader features,  and  it also  leads to  the
emergence of  normally forbidden transitions that  become very intense
at high  field strengths and  that dominate the multiplet.  The higher
Balmer lines  originate from higher angular momentum  l-states and are
therefore subject to greater  Zeeman splitting, i.e., broadening, both
from the normal and quadratic effect.

\begin{figure*}[!p]
\centering
  \includegraphics[angle=-90,width=12cm]{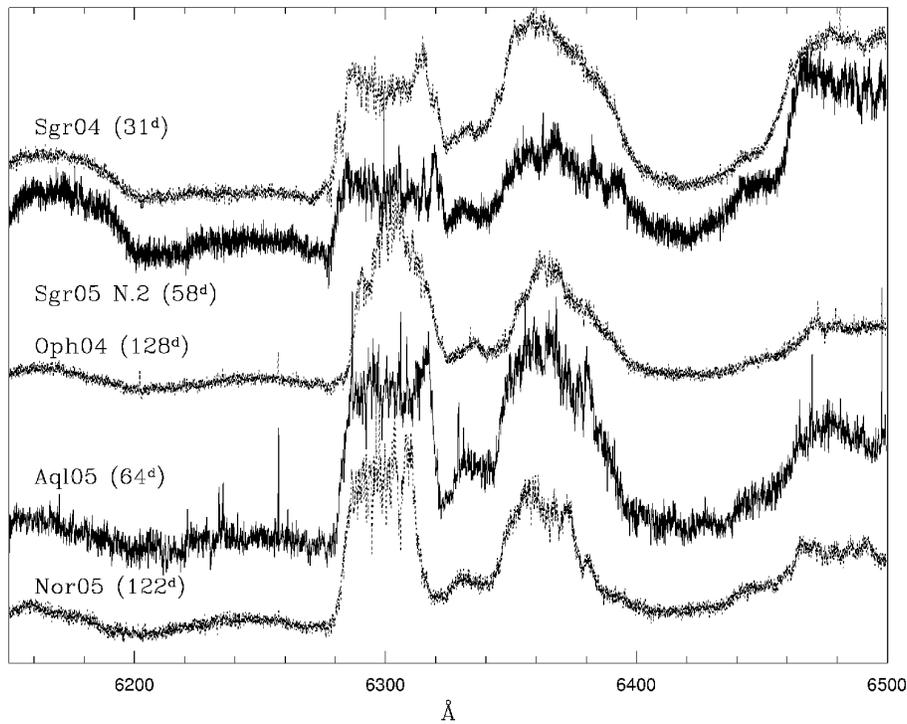}
    \caption{FEROS spectra of five novae showing large differences in
 the profiles of the  [O~I] $\lambda\lambda$ 6300, 6364 lines.  Note the
 emission feature near $\lambda$6330 in each of the spectra.}
  \label{fig:oi}
\end{figure*}

In  normal, low  field  strength conditions  [O~I] $\lambda$6300  and
$\lambda$6364  originate  from  the  same  upper  level  and  have  an
intensity ratio  of 3:1, and  they have identical line  profiles.  The
QZE  alters  this  by  modifying transition  strengths  and  producing
additional  multiplet  components,  such  as the  normally  completely
LS-forbidden $^1S_0  - ^3P_2$  transition. One can  see from  the [O~I]
profiles  shown  in  Fig.~\ref{fig:oi}  that an  emission  feature,  or
component,  appears  near  wavelengths  $\lambda$6330 in  all  of  the
spectra  that  show  the  [O~I]  emission. This  could  be  a  normally
forbidden  transition that emerges  only in  a strong  magnetic field.
The  fact that  this same  behavior is  not observed  for the  [O~III]
$\lambda\lambda$4959,5007 doublet might be attributed to the fact that
the QZE is  most pronounced in neutral atoms  (Jenkins \& Segre 1939),
and [O~III] originates in a region of lower field strength.

Quantitative calculations need to  be carried out to determine whether
the QZE can explain the distinctive aspects of novae spectra described
here because the few existing calculations  of the QZE apply only to H
and He~I.  Ab initio, one  would expect the more  highly ionized gas,
not [O~I], to occupy the regions  close to the white  dwarf where the
magnetic field is the strongest. There is one clear observational test
of  the  Zeeman effect  in  novae, and  that  is  that lines  strongly
influenced by the  QZE, as we suggest for the Balmer  lines and [O~I],
should  be  circularly polarized  (Preston  1970;  Pavlov \&  Shibanov
1976).  A rapidly rotating,  inhomogeneous magnetic field will lead to
global cancellation between the varied local polarization vectors, but
one  would still  expect some  novae to  exhibit  substantial circular
polarization in these transitions.

In  fact, spectrapolarimetry observations  using broad  passbands have
been  made for  a handful  of novae  in the  past decade,  looking for
linear  polarization which has  been detected  at the  $\sim$1\% level
(Johnson et~al.\ 1997). The observed polarization  has been attributed
to  electron scattering  and dust  from  an asymmetric  shell, but  an
alternative  explanation   is  that   it  could  be   residual  linear
polarization produced by the QZE.  Spectrapolarimetric observations of
both linear  and circular polarization  in the Balmer  lines following
outburst, carried out on  large telescopes at moderately high spectral
resolution to  isolate the lines,  should be made as  an observational
test to determine to what extent the Zeeman effect, and especially the
QZE, influences novae spectra.
\begin{table}
\begin{center}
\scriptsize
\caption{}
\begin{tabular}{lccc}
OBJECT & EPOCH & H$\delta$ & H$\beta$ \\ 
Sgr05 N. 1 & \phn48 &  3500 & 2600 \\
          & \phn80 &  3041 & 2640 \\
Sgr05 N. 2 & \phn\phn3 &  1754 & 1893 \\
           & \phn58 &  2150 & 1875 \\
           & \phn81 &  2045 & 1868 \\
Oph04  & 128 &  1880 & 1517 \\
Sco04 N. 2 & \phn43 &  3743 & 3256 \\
Ser05  & \phn88 &  1828 & 1455 \\
\end{tabular}
\end{center}
\end{table}

\section{Post-outburst geometry}

\begin{figure}
  \includegraphics[angle=-90,width=8cm]{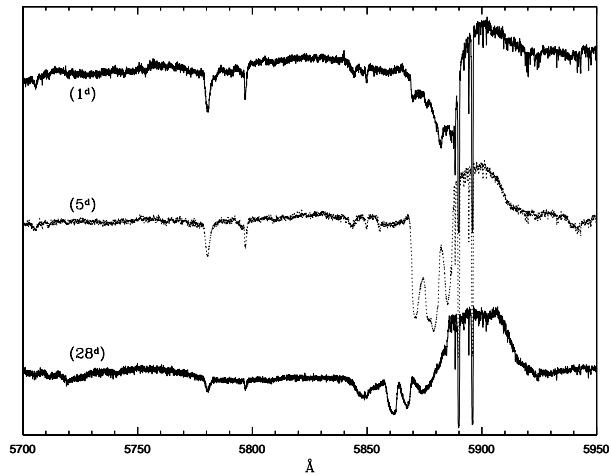}
    \caption{The  evolution  of  Na~I~D  absorption  in Nova  Ophiuchi
  2004. Note the progressive blueward migration of the stronger doublet
   components--a sign of outwardly accelerating gas.}
  \label{fig:na}
\end{figure}

The  geometry of  the novae  ejecta can  be pieced  together  from the
 behavior  of  the  postoutburst  spectra, and  the  absorption  lines
 provide  key information.   One of  the most  common features  of the
 spectra near maximum light is the  strength of the Na~I~D absorption,
 and that  of the Fe~II.  There  are often multiple  components and the
 absorption features  typically migrate to  shorter wavelengths, i.e.,
 increasing  radial velocities,  and  weaken over  timescales of  days
 before  disappearing. Fig.~\ref{fig:na} shows  this behavior  for Nova
 V2574  Oph/04. Given  the large  explosive energy  production  of the
 outburst  the presence  and  persistence of  Na$^0$  (and Fe$^+$)  is
 unexpected  because  it  is  so  easily destroyed  by  ionization  in
 energetic situations due to its  low ionization potential of 5eV. The
 most plausible  source for this  absorbing gas is the  cool secondary
 star in the nova system with perhaps some contribution from the outer
 accretion disk.

 There is  always present a  high velocity postoutburst wind  from the
 white  dwarf of  $>$500 km/s,  evident from  both the  broad emission
 lines  and the broad  absorption normally  seen in  higher ionization
 species such  as He~I $\lambda$5876 that typically  persists for weeks
 to  months following the  outburst.  The  wind impacts  the accretion
 disk and the  secondary star and almost certainly  dislodges gas from
 the outer layers  of the Roche-lobe filling secondary,  which is only
 weakly bound to the star. Such situations can give rise to a momentum
 transfer 'snow  plow' phase with  little energy transfer,  similar to
 that which  occurs in  blast waves where  gas is accelerated  to high
 velocities  while  maintaining  a  low internal  temperature  through
 radiation (Ostriker \& McKee 1988).

Based on the  above considerations the general geometry  of the ejecta
producing  the   postoutburst  spectra   of  novae  consists   of  two
components: a low density,  high velocity, ionized wind originating on
the white dwarf,  and higher density, lower velocity  filaments of gas
that  have been  stripped off  the  secondary star  and the  disrupted
accretion disk, and which are accelerated to high velocities by the WD
wind. The two components mix  with each other, with the higher density
material that originates in the  star and disk being more concentrated
in the  plane of the  binary orbit and  tending to dissipate as  it is
accelerated  by  the  wind.    The  observed  absorption  spectrum  it
therefore dependent on the inclination of the system. The emitting gas
consists of a mixture of nuclear processed, high velocity gas from the
white dwarf and unprocessed material from the secondary star and disk,
and the  interpretation of the  abundances of the emitting  gas should
take into account the mixture of the two components.

Series of  high resolution spectra covering a  broad wavelength regime
and  obtained at  different epochs  following the  outburst  present a
detailed   picture  of   postoutburst  novae.   We  are   preparing  a
presentation  of  the  data  from  the  FEROS  survey  in  a  separate
publication, including a  more detailed description and interpretation
of the spectra.

\end{document}